\documentclass[aps,prl,twocolumn,showpacs,floatfix,nofootinbib]{revtex4}
\usepackage{graphicx}
\usepackage{hyperref}
\usepackage{amsmath}
\usepackage{amssymb}

\def\be{\begin{equation}}
\def\ee{\end{equation}}
\def\bea{\begin{eqnarray}}
\def\eea{\end{eqnarray}}
\def\>{\rangle}
\def\<{\langle}

\newcommand{\qed}{\hfill$\square$\par\vskip12pt}

\begin{document}
\title{A relational quantum computer using only two-qubit total spin measurement and an initial
supply of highly mixed single qubit states.}
\author{Terry Rudolph \footnote{email: t.tudolph at imperial.ac.uk}}
\author{Shashank Soyuz Virmani \footnote{email: s.virmani at imperial.ac.uk}}
\affiliation{QOLS, Blackett Laboratory, Imperial College, Prince
Consort Road, London SW7 2BW, UK}

\begin{abstract}
We prove that universal quantum computation is possible using only
(i) the physically natural measurement on two qubits which
distinguishes the singlet from the triplet subspace, and (ii)
qubits prepared in almost any three different (potentially highly
mixed) states. In some sense this measurement is a ``more
universal'' dynamical element than a universal 2-qubit unitary
gate, since the latter must be supplemented by measurement.
Because of the rotational invariance of the measurement used, our
scheme is robust to collective decoherence in a manner very
different to previous proposals - in effect it is only ever
sensitive to the relational properties of the qubits.
\end{abstract}

\maketitle

There has been considerable effort directed toward understanding
how measurements can be utilized in quantum computation
\cite{everyone,briegelcluster} - either as part of error
correction, or as a method of replacing some (or all) of the
coherent processes of the standard unitary circuit model. With
respect to this latter program, many proofs of universality are
for abstract measurements - often on more than two systems - and
little attention has been focussed on measurements that are
physically natural.

An example of a physically natural measurement, is the parity
measurement on generic bosonic systems - a measurement which
compares the state of two systems as to whether they are the
``same'' or ``different''. Parity measurements can be performed
using Clifford operations \cite{cliff}, and have formed some part
of various measurement based schemes. Recently it was shown that
certain non-deterministic parity measurements, along with single
qubit unitaries, are universal for quantum computation
\cite{danterry}, and the proof allows for a dramatic
simplification of the resources required for linear optical
quantum computation.

In this paper we will focus on a different physically natural
two-qubit measurement that is not a Clifford operation
 (we will see why this is the case later). Abstractly,
the measurement is composed of the projectors \be J_0\equiv
|\psi^-\>\<\psi^-|,\;\;\;J_1 \equiv I-J_0, \ee where $|\psi^-\>$
is the singlet state. As a measurement on two spin-1/2 systems,
this rotationally invariant ``$J$-measurement'' is one of total
angular momentum - a projection onto the singlet or triplet states
according to whether the total angular momentum is 0 or 1
respectively. $J$-measurements are physically natural primarily
because in a wide variety of atomic and solid state systems the
natural interaction Hamiltonians have different energies for the
singlet versus the triplet states.

Various results on the universality of $J$-measurements for
quantum computation can be readily obtained. For instance,
 it can be shown that $J$-measurements, single qubit
measurements \& unitaries (in particular, say, just Hadamard and
phase gates), and systems initialized in the computational basis
state $|0\>$, are universal. In fact, it can even be shown that
\emph{any} two outcome measurement composed of projectors
$\{|\phi\>\<\phi|,I-|\phi\>\<\phi|\}$, for an arbitrary two-qubit
state $|\phi\>$, is universal under similar conditions.

However, here we will show the much stronger result that
\begin{quote}\textbf{Theorem 1:}\textit{
 Quantum computation can be performed using (a) two
qubit $J$-measurements, and (b) \emph{any} (polynomially large)
supply of single qubit mixed states prepared along three linearly
independent Bloch vectors.}
\end{quote}
Our scheme has a number of interesting features:
\begin{itemize}
\item We do not require single-qubit measurements, nor do we require the initial
supply of states to be pure. In fact, the initial qubit states can
be very highly mixed, as long as none of them are maximally mixed.
To date, all other schemes either require more than one type of
measurement and/or require initially pure states and/or require
joint measurements on more than two qubits.

\item The $J$-measurement is the only dynamical object used in the computation.
It is therefore ``more universal'' than a universal two-qubit
unitary operation, which must be supplemented by some form of
measurement.

\item The $J$-measurement is rotationally invariant - the
measurement is sensitive only to the relative state of the systems
involved (in fact, it is the optimal measurement for determining
relative information of two qubits \cite{BRS2}). As such, the
computation is naturally robust to random collective rotations -
not because the states are rotationally invariant, as is the
standard procedure for protecting against such decoherence, but
rather because the dynamics is robust.
\end{itemize}

We will prove Theorem 1 utilizing the cluster-state model of
quantum computation \cite{briegelcluster}. We focus on the
simplest path to proving universality - for many of our
constructions there exist more efficient, but less transparent,
procedures. We will heavily use the following three useful
properties of the $J$-measurement, proofs of which will be given
later:

\smallskip

\noindent\textbf{Property (i)} \textit{Purification:} Given a
supply of a single qubit mixed state $\rho$ with Bloch vector
$r_0\hat{r}$, $r_0 \geq 0$, and the capacity to do
$J$-measurements, it is possible to prepare single qubit states
with Bloch vector $(1-\epsilon)\hat{r}$ using constant resources
for any fixed $\epsilon > 0$. (The fact that such purification can
be achieved is an easy way to see that the J-measurement is not a
Clifford operation - it is known that Clifford operations cannot
be used to purify arbitrary Bloch vectors \cite{BK}.)



\smallskip

\noindent\textbf{Property (ii)} \textit{Programmable single qubit
measurements:} Given $2^n-1$ copies of a pure single qubit state
$|\phi\>$, a projective measurement of an arbitrary qubit onto the
orthogonal pair of states $\{|\phi\>,|\bar{\phi}\>\}$ can be
simulated by $J$-measurements, with fidelity that goes as
$1-1/2^n$. (This measurement does not collapse the qubit being
measured in the same way as a standard projective measurement
would, however it (asymptotically) collapses a remote system with
which the measured qubit is entangled in the standard way, and
this suffices for our purposes.)


\smallskip

\noindent\textbf{Property (iii)} \textit{Creation of maximally
entangled states:} It is possible to create all four Bell states
($|\psi^{\pm}\rangle := {1 \over \sqrt{2}}(|01\> \pm |10\>),
|\phi^{\pm}\rangle := {1 \over \sqrt{2}}(|00\> \pm |11\>)$), as
well as all GHZ states of the form: $|0\>^{\otimes
n}\pm|1\>^{\otimes n}$.

Using the above properties, we now prove the theorem.

\noindent\textbf{Proof of Theorem 1:}

From Properties (i) and (ii) it is clear that the single qubit
states prepared, and the single qubit measurements implemented are
only asymptotically ``sharp''. However, in \cite{dawson}
fault-tolerant procedures for cluster state computation have been
demonstrated for certain types of noise, including situations in
which the cluster state is prepared non-deterministically. The
types or error incurred by our scheme's inherent imperfections
fall into the class of error models considered by \cite{dawson},
with at most minor restrictions on the topology of the cluster
states that we are allowed to make. Hence there exists a finite
fault tolerant threshold that we need to obtain. This threshold
can be obtained with constant effort, and hence for now we will we
proceed as if properties (i)-(iii) involve no imperfections
whatsoever. Given this assumption, the proof proceeds via a
sequence of primitive operations, and so we will first demonstrate
how each of these primitives may  be achieved. In the rest of the
paper we will frequently omit normalisation factors from our
equations in order not to clutter the notation.

\medskip

\noindent {\it Creating ``flipped'' qubit states.} From a qubit
mixed state with Bloch vector $\vec{r}$ we can create a
``spin-flipped'' qubit with Bloch vector $-\vec{r}/3$ using the
following procedure (an optimal cloning/spin flipping): Take an
ancillary pair of qubits in a singlet state (which is readily
obtained), and implement the $J$-measurement on one member of the
singlet and the qubit to be flipped \cite{what}. If the $J_1$
outcome is obtained, which occurs with probability 3/4, it is
easily verified the reduced density matrix of the unmeasured
singlet qubit now has Bloch vector $-\vec{r}/3$.

\smallskip

\noindent {\it Creating arbitrary qubit pure states.} By flipping
each of our original set of 3 types of qubits with linearly
independent Bloch vectors, we can efficiently create qubits with 6
different Bloch vectors. Probabilistically mixing these states
allows us to generate any mixed state that lies inside the
polyhedron that has these 6 states as its vertices. Simple
geometrical considerations show that such a polyhedron necessarily
contains a sphere of finite radius centered at the origin. Any
such states can then be purified (using Property (i)), leading to
the creation of \emph{arbitrary} pure states.


\smallskip

\noindent {\it Creating 2-qubit cluster states.} By Property (iii)
we can create all four Bell states. We can now create the
maximally entangled two qubit cluster state $|0+\>+|1-\>$ as
follows. Take four qubits in the state
$|\psi^-\>_{12}\otimes|\phi^+\>_{34}$ and perform a
$J$-measurement between qubit pairs 1,3 and 2,4. In the event of
obtaining the $J_1$ outcome on both measurements (which occurs
with probability 1/2), the four qubits are collapsed to the state
$|\phi^-\>_{14}|\psi^+\>_{23}-|\psi^+\>_{14}|\phi^-\>_{23}$. It is
readily verified that now performing a single qubit measurement on
qubit 1 in the $|0\>,|1\>$ basis, and qubit 4 in the $|\pm\>$
basis, collapses qubits 2,3 to a two qubit cluster state,
regardless of the outcome\footnote{We are considering cluster
states which differ only by Pauli operations as equivalent, since
such differences can be compensated for classically, in the
standard manner, during the course of the cluster computation.}.

\smallskip

\noindent {\it Redundant encoding of cluster states.} In order to
create larger cluster states, we will need to utilize the concept
of a \emph{redundant encoding} of a given qubit in the cluster
\cite{danterry}. Such an encoding is one in which extra physical
qubits are appended ``in parallel'', such that they are still
considered to be part of the logical encoding of a single cluster
state qubit. More precisely: a generic cluster state can be
written $(|X\>|0\>+|X^\perp\>|1\>)$, where we have singled out one
qubit from the cluster state. A 4-qubit redundant encoding of this
qubit would be $(|X\>|0\>|000\>+|X^\perp\>|1\>|111\>)\equiv
(|X\>|0^4\>+|X^\perp\>|1^4\>)$. Note that it is not necessary for
the states in a redundant encoding to be the ``same'' qubit state
in parallel - the two redundant strings need only be orthogonal at
each position, and in particular could be any bitwise orthogonal
strings of \{0,1\}. We will soon show how such a redundant
encoding may be achieved. At the point of performing a cluster
computation, unwanted redundantly encoded qubits can be removed by
measurement in the $|\pm\>$ basis - for example, measuring and
then discarding the last qubit in the redundantly encoded cluster
$(|X\>|0\>|0\>+|X^\perp\>|1\>|1\>)$ gives one of $|X\>|0\> \pm
|X^\perp\>|1\>$, which are the same as the unencoded cluster state
up to at most an unimportant Z rotation on one qubit.

\smallskip

\noindent {\it Fusion of small clusters into larger ones.}
Consider two independent cluster states, with one qubit singled
out from each, such that the states can be written
\[(|X\>|0\>+|X^\perp\>|1\>)\otimes(|Y\>|0\>+|Y^\perp\>|1\>).\] A fusion operation
produces one of the states
\[
(|X\>|Y\>|0\>+|X^\perp\>|Y^\perp\>|1\>),(|X\>|Y^\perp\>|0\>+|X^\perp\>|Y\>|1\>),
\]
or any states obtained from these via a Pauli gate applied to the
singled out cluster qubit. It can readily be verified that these
two states are essentially equivalent larger cluster states formed
by fusing the original smaller clusters, as depicted graphically
in Fig.~1. The fusion operation may be achieved as follows.
Imagine that we have created two independent clusters that have
been redundantly encoded. Suppose that we wish to fuse two of the
qubits, one from each cluster. Suppose that these two qubits are
encoded using numbers $a$ and $b$ of qubits respectively.
Implementing a $J$-measurement between one member of each cluster
qubit's redundant encoding yields the following two possible
evolutions, where we have reordered the states so that the
measured qubits appear at the very end of each term:
 \begin{eqnarray*}
&&(|X\>|0^a\>+|X^\perp\>|1^a\>)\otimes (|Y\>|0^b\>+|Y^\perp\>|1^b\>)\Rightarrow\\
J_0:&& \left(|X\>|Y^\perp\>|0^{a-1}\>|1^{b-1}\>-|X^\perp\>|Y\>|1^{a-1}\>|0^{b-1}\>\right)|\psi^-\>\\
J_1:&& |X\>|Y\>|0^{a+b}\>+|X^\perp\>|Y^\perp\>|1^{a+b}\>+\\
&&(|X\>|Y^\perp\>|0^{a-1}\>|1^{b-1}\>+|X^\perp\>|Y\>|1^{a-1}\>|0^{b-1}\>)\frac{|\psi^+\>}{\sqrt{2}}\\
\end{eqnarray*}
The $J_0$ outcome (which occurs with probability 1/4) is fine -
after throwing away the residual singlet, it amounts to having
simply fused the two qubits into a new cluster qubit with a
redundant encoding of $a+b-2$ qubits. However, the second term in
the expression resulting from the $J_1$ outcome is undesirable. To
project out this piece of the state, we note that this piece has
the two qubits which were measured in the $|\psi^+\>$ state,
whereas the desired piece has these two qubits in either $|00\>$
or $|11\>$. Note also that $|\psi^+\>=|++\>-|--\>$. Thus if single
qubit measurements are performed in the $|\pm\>$ basis on these
two qubits, and anticorrelated outcomes are obtained (i.e. $|+-\>$
or $|-+\>$) then the $|\psi^+\>$ part is projected out, and  the
desired fused cluster state is obtained. The overall joint
probability of obtaining the $J_1$ outcome followed by these
anticorrelated outcomes is 1/4.

A failure occurs when the correlated outcome ($|++\>$ or $|--\>$)
is obtained - the overall probability of such a failure is 1/2.
Crucially, however, when a failure \emph{does} occur it is easy to
see that the final state of the two clusters is one of
\[
(|X\>|0^{a-1}\>\pm|X^\perp\>|1^{a-1}\>)\otimes
(|Y\>|0^{b-1}\>\pm|Y^\perp\>|1^{b-1}\>),
\]
depending upon the outcome. Thus we see that the original cluster
states have not been destroyed, all that has happened is one qubit
has been removed from each cluster qubit's redundant encoding. If
sufficient qubits remain, the fusion can be reattempted.


Clearly if we start off with two-qubit cluster states with large
enough redundant encodings, then they can be fused with high
probability. As in \cite{danterry,nielsen}, simple random walk
considerations then show that we efficiently create clusters of
arbitrary size by such probabilistic fusion.

We have seen above that we can create a two qubit cluster
$|0+\>+|1-\>$, which has no redundant encoding - what remains,
therefore, is to verify that we can create such clusters with a
suitable amount of redundant encoding. If we consider a two qubit
cluster and a GHZ state:
\[(|0+\>+|1-\>)\otimes (|0^a\>+|1^a\>).
\]
we readily verify that applying the fusion procedure outlined above
simply creates a suitable redundant encoding. Unsuccessful such
fusions need to be discarded - since we envisage such a procedure is
being implemented ``offline'' this is not an issue. (Similarly, such
fusion can be used to probabilistically create large GHZ states from
smaller ones.) \qed

Note that Theorem 1 leaves open the interesting question whether
we actually need three linearly independent supplies of qubit
states. It is possible that via a smarter encoding than these
authors are capable of finding, it might be possible to perform
quantum computation with $J$-measurements and only {\it one} or
{\it two} different types of initial state.

\begin{figure}
\includegraphics[scale=0.3]{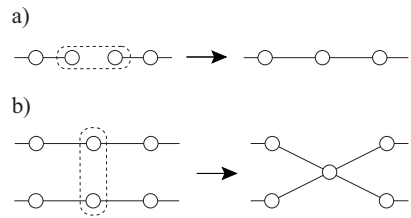}
\caption{\label{fig:fuse} Examples of how clusters states can be
joined together by ``fusing''  qubits from each.}
\end{figure}

\textbf{Proof of Property (i):} Provably optimal purification to
the largest eigenvector of $\rho$ by using large joint
measurements of total angular momentum was described in
\cite{cirac}. Here we perform purification that is not optimal in
terms of resources, but uses only the two-qubit $J$-measurement.

Assume that we have $N$ copies of $\rho$ whose Bloch vector is
$\vec{r}_0.$ Sort these into $N/2$ pairs and perform the
$J$-measurement, keeping only those pairs for which $J_1$ is
obtained. The Bloch vector of the reduced state on either side is
now $\vec{r}_1=\frac{4}{3+r_0^{2}}\vec{r}_0.$ The probability of
getting the $J_{1}$ outcome is simply
$P(r_0)=\frac{3+r_0^{2}}{4}.$ Imagine we throw away half of these
pairs. We are
left with $\frac{NP(r_0)}{2}$ systems which have the longer Bloch vector $%
\vec{r}_1.$

How many times must we repeat the process so that the Bloch vector
has length $r^{\max }\geq 1-\varepsilon $ ? Multiple repetitions
of the above process yields the recursion relation $
r_{n+1}=4r_n/(3+r_n^{2}) $ This is not easily solved. However, we
underestimate the growth of the Bloch vector if we presume it
follows the simpler recurrence $ R_{n+1}=4R_n/(3+R_n),$ that is,
$r_n\ge R_n$ $\forall n$, with $R_0=r_0$. This latter recurrence
{\it is} easily solved, yielding
\[
R_n=\left(\left( \frac{3}{4}%
\right) ^{n}\frac{\left( 1-r_{0}\right) }{r_{0}}+1\right)^{-1}
\]
From this we deduce that to obtain $R_n\geq 1-\epsilon$ it suffices
to take
\[
n \geq \log \left( \frac{(1-\epsilon )(1-r_{0})}{\epsilon r_{0}}%
\right) /\log \frac{4}{3},
\]
or, more simply, $n\ge 3\log \left( \frac{1}{\epsilon
r_{0}}\right)$ will suffice. Obviously we will not always be
successful in obtaining the $J_1$ outcome, and so this process
will only succeed with some probability. However, as we only need
to obtain some constant fault tolerance threshold, the resource
requirements for this purification procedure are still constant,
and the above arguments are sufficient to show that our
purification procedure is efficient enough. Nevertheless, for
completeness we may perform an approximate analysis of the
overheads involved in our purification method. The probability of
success (i.e. obtaining the $J_1$ outcome) on any given pair is
$P(r_n)=(3+r_n^{2})/4.$ Thus the fraction $\eta$ of the original
$N$ qubits which have been successfully purified to length
$1-\epsilon$ after $m= 3\log \left( \frac{1}{\epsilon
r_{0}}\right)$ steps  is
\[ \eta=\frac{1}{2^m}\prod_{j=0}^{m-1} P(r_j)\ge (r_0\epsilon)^3 \prod_{j=0}^{m-1} P(r_j)  .
\]
(The factor of $1/2^m$ arises from the discarding of half the
successful qubits at every step; we take $3\log \left( 1/\epsilon
r_{0}\right)$ to be an integer.) Since $P(r_j)\ge
(3+R_j)/4=R_j/R_{j+1}$, we easily lower bound the product of
probabilities to obtain $\eta \ge (r_0\epsilon)^3
\frac{R_0}{R_m}\ge (r_0\epsilon)^3 r_0.$\qed

\textbf{Proof of Property (ii):} Given a supply of qubits in a
state $|\phi\>$, we will show how to effect a destructive
measurement in the basis $|\phi\>,|\phi^{\perp}\>$ with an
arbitrarily small inaccuracy. By the rotational invariance of the
$J$-measurement, we can assume that this basis is actually the
computational basis. Hence suppose that we have $2^n-1$ ancilla
qubits prepared in the state $|0\>$, labeled $2,3,\ldots 2^n$, and
that we wish to approximates a destructive measurement
$\{|0\>\<0|,|1\>\<1|\}$ of a single input qubit, labeled qubit 1.
We consider the following measurement strategy: Measure qubits
1\&2. If $J_0$ is obtained, the measurement outcome is declared
$|1\>\<1|$. If $J_1$ is obtained, measure qubit pairs ($1,3$) and
($2,4$). If $J_0$ is obtained on either pair, the measurement
outcome is declared $|1\>\<1|$. If $J_1$ is obtained on both
pairs, we measure pairs $(1,5),(2,6),\ldots,(4,8)$. We continue in
this fashion, until either we obtain a $J_0$ outcome, or, in the
final step, we obtain all $J_1$ outcomes on the measurements of
qubit pairs $1\&(2^{n-1}+1),\ldots,2^{n-1}\&2^n$. In this latter
case we declare the measurement outcome to be $|0\>\<0|$.

Clearly, if the input state was $|0\>$, the qubits are in a
symmetric state and only the $J_1$ outcome will ever be obtained.
In order to understand the more complicated case when the input
qubit is in state $|1\>$, it helps to note that the operator $J_1$
acting on qubits $i,j$ can be written
$J_1^{ij}=\frac{1}{2}\left(I+F_{ij}\right)$, where $F_{ij}$ is the
unitary ``SWAP'' operation, which swaps qubits $i$ and $j$. An
error occurs in the above measurement procedure whenever only the
$J_1$ outcome is always obtained, despite the input state being 1.
If we define $|I_k\>$ to be an equiweighted superposition of the
$k$ states with hamming weight 1 ($|I_2\>=|\psi^+\>$), it can be
readily seen that when an error occurs the input state evolves as
follows:
\[
|1\>|0\>^{\otimes L}\longrightarrow |I_2\>|0\>^{\otimes 2^n-2}
                   \longrightarrow|I_4\>|0\>^{\otimes 2^n-3}
                    \ldots
                    \longrightarrow |I_{2^n}\>.
\]
It is easy to verify the probability of such an undesired
evolution occurring is simply $1/2^n$. This whole process is
effectively a form of `programming' of quantum measurements
\cite{prog}, with the difference that we are using only a
two-qubit `detector'.\qed

\textbf{Proof of Property (iii):} Consider performing the
$J$-measurement on qubits initially in the state $|01\>$. With
equal likelihood the qubits are collapsed into the $|\psi^-\>$ and
$|\psi^+\>$ states. If a $J_1$ outcome is obtained after
measurement on a pair of qubits initially in the states
$|+\>|-\>$, then they are collapsed into the state $|\phi^-\>$.
Finally, if a $J_1$ outcome is obtained after measurement on a
pair of qubits initially in the states
$(|0\>+i|1\>)\otimes(|0\>-i|1\>)$, then the qubits are collapsed
into the $|\phi^+\>$ state. Thus we can create all 4 Bell states.

To create a GHZ state, consider taking four qubits, initially in
the states $|\phi^+\>_{12}\otimes |\phi^-\>_{34}$ and performing
$J$-measurements on the pairs $(1,3),(2,4)$. In the event of $J_1$
outcomes being obtained on both pairs, the four qubits are
collapsed into the state $|0000\>-|1111\>$. This state suffices
for our purposes.\qed


%

\vspace{1cm}

\begin{acknowledgments}
We thank Dan Browne, Artur Ekert, Martin Plenio and Tom Stace for
interesting discussions. We also thank Thomas P. Stafford, Vance
D. Brand, Donald K. `Deke' Slayton, Alexei Leonov, and Valeri
Kubasov, whose historic meeting in space inspired Mummy and Daddy
Virmani in the naming of their son. T.R. also thanks Mummy and
Daddy Virmani for not having chosen the other obvious middle name.
We acknowledge funding from the EPSRC and the Royal Commission for
the Exhibition of 1851.

\end{acknowledgments}

\end{document}